\documentclass[12pt]{article}%
\usepackage{graphicx}
\usepackage{amsmath}
\usepackage{epsfig}
\usepackage{amsfonts}
\usepackage{amssymb}%
\providecommand{\U}[1]{\protect\rule{.1in}{.1in}}
\begin{document}

\title{Pair Photoproduction in Constant and Homogeneous Electromagnetic Fields }
\author{V.M. Katkov\\Budker Institute of Nuclear Physics,\\Novosibirsk, 630090, Russia\\e-mail: katkov@inp.nsk.su}
\maketitle

\begin{abstract}
The process of pair creation by a photon in a constant and homogeneous
electromagnetic field of an arbitrary configuration is investigating. At high
energy the correction to the standard quasiclassical approximation (SQA) has
been calculated. In the region of intermediate photon energies where SQA is
inapplicable the new approximation, developed recently by authors, is used.
The influence of weak electric field on the process in a magnetic field is
considered. In particular, in the presence of electric field the root
divergence in the probability of pair creation on the Landau energy levels is
vanished. For smaller photon energies the low energy approximation is used.
The found probability describes the absorption of soft photon by particles
created by field. At low photon energy the electric field action dominates and
the influence of magnetic field on the process is connected with the
interaction of it and the magnetic moment of creating particles.

\end{abstract}

\section{\ \ Introduction}

The pair photoproduction in an electromagnetic field is the basic QED reaction
which can play the significant role in many processes.

This process was considered first in a magnetic field. The investigation was
started in 1952 independently by Klepikov and Toll \cite{[1],[2]}. In
Klepikov's paper \cite{[3]}, which was based on the solution of the Dirac
equation, the probability of photoproduction had been obtained on the mass
shell ( $k^{2}=0,k$ is the 4-momentum of photon. We use the system of units
with $\hbar=c=1$ and the metric $ab=a^{\mu}b_{\mu}=a^{0}b^{0}-\boldsymbol{ab}%
$). In 1971 Adler \cite{[4]} had calculated the photon polarization operator
in a magnetic field using the proper-time technique developed by Schwinger
\cite{[5]} and Batalin and Shabad \cite{[6]} had calculated this operator in
an electromagnetic field using the Green function found by Schwinger
\cite{[5]}. In 1975 the contribution of charged-particles loop in an
electromagnetic field with $n$ external photon lines had been calculated in
\cite{[7]}. For $n=2$ the explicit expressions for the contribution of scalar
and spinor particles to the polarization operator of photon were given in this
work. Making use of the imaginary part of this operator for spinor particles
the pair photoproduction probability was analyzed in the pure magnetic
\cite{[8]} and the pure electric \cite{[9]} field.

The probability of pair photoproduction in a constant and homogeneous electric
field in the quasiclassical approximation had been found by Narozhny
\cite{[10]} using the solution of the Dirac equation in the Sauter potential
\cite{[11]}. Nikishov \cite{[12]} had obtained the differential distribution
of this process also using the solution of the Dirac equation in the indicated field.

In the present paper we consider the integral probability of pair creation in
a constant and homogeneous electromagnetic field of an arbitrary configuration
basing on the polarization operator \cite{[7]}. In Sec.2 the exact expression
for this probability has been received for the general case $k^{2}\neq0.$In
Sec.3 the standard quasiclassical approximation (SQA) is outlined for the
high-energy photon $\omega\gg m$ ( $m$ is the electron mass). The corrections
to SQA, determined also the applicability region of SQA, have been calculated.
The found expressions, given in the Lorentz invariant form, contain two
invariant parameters. In Sec.4 the new approach has been developed for the
relatively low energies where SQA is not applicable. This approach is based on
the method proposed in \cite{[8]}. The obtained probability is valid in the
wide interval of photon energy, which is overlapped with SQA. In Sec.5 the
case of the "nonrelativistic" photon $\omega\ll m$ is analyzed. In particular,
in the energy region $\omega\lesssim$ $eE/m$ where the previous approach is
inapplicable, the low energy approximation has been developed basing on the
analysis in \cite{[9]}. In tern the found results have an overlapping region
of applicability with the previous approach. So just as in \cite{[9]} we have
three overlapping approximations which include all photon energies. At the
photon energy $\omega\ll$ $eE/m$ the probability has been found for arbitrary
values of fields $E$ and $H.$

\section{\ \ General expressions for the probability of process}

Our analysis is based on the general expression for the contribution of spinor
particles to the polarization operator obtained in a diagonal form in
\cite{[7]} (see Eqs. (3.19), (3.33)). The imaginary part of the eigenvalue
$\kappa_{i}$ of this operator on the mass shell $(k^{2}=0)$ determines the
probability per unit length $W_{i}$ of $e^{-}e^{+}$ pair creation by the real
photon with the polarization $e_{i}$ directed along the corresponding
eigenvector:%
\begin{equation}
W_{i}=-\frac{\mathrm{\operatorname{Im}}\kappa_{i}}{\omega};\ \ e_{i}^{\mu
}=\frac{b_{i}^{\mu}}{\sqrt{-b_{i}^{2}}},~\ b_{2}^{\mu}=\ \left(  Bk\right)
^{\mu}+\frac{2\Omega_{4}}{\Omega}\left(  Ck\right)  ^{\mu},\ \label{eq1}%
\end{equation}

\[
\ b_{3}^{\mu}=\left(  Ck\right)  ^{\mu}-\frac{2\Omega_{4}}{\Omega}\left(
Bk\right)  ^{\mu};
\]

\begin{align}
-\mathrm{\operatorname{Im}}\kappa_{2}  &  =r\left(  \Omega_{2}-\frac
{2\Omega_{4}^{2}}{\Omega}\right)  ,\ \ \ -\mathrm{\operatorname{Im}}%
\ \kappa_{3}=r\left(  \Omega_{3}+\frac{2\Omega_{4}^{2}}{\Omega}\right)
,\ \ \label{eq2}\\
\ \Omega &  =\Omega_{3}-\Omega_{2}+\sqrt{(\Omega_{3}-\Omega_{2})^{2}%
+4\Omega_{4}^{2}},\ \ r=\frac{\omega^{2}-k_{3}^{2}}{4m^{2}}.\nonumber
\end{align}
The consideration realizes in the frame where electric $\mathbf{E}$ and
magnetic\textbf{ }$\mathbf{H}$ fields are parallel and directed along the axis
3. In this frame the tensor of electromagnetic field $F_{\mu\nu}$ and tensors
$F_{\mu\nu}^{\ast}$, $B_{\mu\nu}$ and $C_{\mu\nu}$ have a form%

\begin{align}
F_{\mu\nu}  &  =C_{\mu\nu}E+B_{\mu\nu}H,\ \ F_{\mu\nu}^{\ast}=C_{\mu\nu
}H-B_{\mu\nu}E,\ \ C_{\mu\nu}=g_{\mu}^{0}g_{\nu}^{3}-g_{\mu}^{3}g_{\nu}%
^{0},\ \nonumber\\
\ B_{\mu\nu}  &  =g_{\mu}^{2}g_{\nu}^{1}-g_{\mu}^{1}g_{\nu}^{2};\ \ eE/m^{2}%
=E/E_{0}\equiv\nu,\ \ eH/m^{2}=H/H_{0}\equiv\mu;\label{eq3}\\
\Omega_{i}  &  =\frac{\alpha m^{2}}{2\pi\text{\textrm{i}}}\mu\nu
\int\limits_{-1}^{1}\ dv\int\limits_{-\infty-\text{\textrm{i}0}}%
^{\infty-\text{\textrm{i}0 }}\ f_{i}(v,x)\exp(\text{\textrm{i}}\psi(v,x))xdx.
\label{eq4}%
\end{align}
Here%

\begin{align}
f_{1}  &  =\frac{\cos(\mu xv)\cosh(\nu xv)}{\sin(\mu x)\sinh(\nu x)}%
-\frac{\cos(\mu x)\cosh(\nu x)\sin(\mu xv)\sinh(\nu xv)}{\sin^{2}(\mu
x)\sinh^{2}(\nu x)},\nonumber\\
f_{2}  &  =2\frac{\cosh(\nu x)(\cos(\mu x)-\cos(\mu xv))}{\sinh(\nu x)\sin
^{3}(\mu x)}+f_{1},\ \nonumber\\
\ \ f_{3}  &  =2\frac{\cos(\mu x)(\cosh(\nu x)-\cosh(\nu xv))}{\sin(\mu
x)\sinh^{3}(\nu x)}-f_{1},\nonumber\\
f_{4}  &  =\frac{\cos(\mu x)\cos(\mu xv)-1}{\sin^{2}(\mu x)}\frac{\cosh(\nu
x)\cosh(\nu xv)-1}{\sinh^{2}(\nu x)}\nonumber\\
&  +\frac{\sin(\mu xv)\sinh(\nu xv)}{\sin(\mu x)\sinh(\nu x)};\label{eq5}\\
\ \ \psi(v,x)  &  =2r\left(  \frac{\cosh(\nu x)-\cosh(\nu xv)}{\nu\sinh(\nu
x)}+\frac{\cos(\mu x)-\cos(\mu xv)}{\mu\sin(\mu x)}\right)  -x. \label{eq6}%
\end{align}
Let us note that the integration contour in Eq.(\ref{eq4}) is passing slightly
below the real axis.

After all calculations have been fulfilled we can return to a covariant form
of the process description using the following expressions%

\begin{align}
E^{2},H^{2}  &  =\left(  \mathcal{F}^{2}+\mathcal{G}^{2}\right)  ^{1/2}%
\pm\mathcal{F,\ \ F=}\left(  \mathbf{E}^{2}-\mathbf{H}^{2}\right)
\diagup2,\ \ \mathcal{G=}\mathbf{EH},\ \nonumber\\
\left(  C^{2}\right)  _{\mu\nu}  &  =\left(  F_{\mu\nu}^{2}+H^{2}g_{\mu\nu
}\right)  \diagup\left(  E^{2}+H^{2}\right)  ,\ \ \left(  C^{2}\right)
_{\mu\nu}-\left(  B^{2}\right)  _{\mu\nu}=g_{\mu\nu}. \label{eq7}%
\end{align}

\section{\ \ Quasiclassical approximation}

The standard quasiclassical approximation (SQA) was developed first for a
magnetic field in \cite{[3]}, \cite{[13]}, \cite{[14]}. The SQA is valid for
ultrarelativistic created particles ( $r\gg1$) and can be derived from
Eqs.(\ref{eq4})-(\ref{eq6}) by expanding the functions $f_{i}(v,x)$,
$\psi(v,x)$ over $x$ powers. To get the correction to the probability in SQA
we shall keep leading to leading powers of $x$. We have%

\begin{equation}
~\ b_{2}^{\mu}=\ \left(  Bk\right)  ^{\mu}+\frac{\nu}{\mu}\left(  Ck\right)
^{\mu}\propto F^{\mu\nu}k_{\nu},\ \ b_{3}^{\mu}=\left(  Ck\right)  ^{\mu
}-\frac{\nu}{\mu}\left(  Bk\right)  ^{\mu}\propto F^{\ast\mu\nu}k_{\nu};
\label{eq8}%
\end{equation}

\begin{align}
-\mathrm{\operatorname{Im}}\kappa_{i}  &  =\mathrm{i}\frac{\alpha m^{2}}%
{12\pi}r(\mu^{2}+\nu^{2})\int\limits_{-1}^{1}dv\left(  1-v^{2}\right)
\int\limits_{-\infty}^{\infty}h_{i}(v,x)\left[  -\mathrm{i}\gamma\left(
v,x\right)  \right]  xdx;\nonumber\\
\gamma\left(  v,x\right)   &  =x+\frac{x^{3}}{12}r\left(  1-v^{2}\right)
^{2}\left(  \nu^{2}+\mu^{2}\right)  ,\label{eq9}\\
h_{2}(v,x)  &  =\frac{3+v^{2}}{2}+\frac{1}{30}\left(  15-6v^{2}-v^{4}\right)
\left(  \mu^{2}-\nu^{2}\right)  x^{2}\nonumber\\
&  -\frac{\mathrm{i}}{720}r(\mu^{2}+\nu^{2})\left(  1-v^{2}\right)
^{2}\left(  9-v^{2}\right)  \left(  \mu^{2}-\nu^{2}\right)  x^{5},\nonumber\\
h_{3}(v,x)  &  =3-v^{2}+\frac{1}{60}\left(  15-2v^{2}+3v^{4}\right)  \left(
\mu^{2}-\nu^{2}\right)  x^{2}\nonumber\\
&  -\frac{\mathrm{i}}{360}r(\mu^{2}+\nu^{2})\left(  1-v^{2}\right)
^{2}\left(  3-v^{2}\right)  ^{2}\left(  \mu^{2}-\nu^{2}\right)  x^{5}.
\label{eq10}%
\end{align}
\newline We are using the known integrals:%

\begin{align}
\int\limits_{-\infty}^{\infty}\cos\left(  x+\frac{ax^{3}}{3}\right)  dx  &
=\frac{2}{\sqrt{3a}}K_{1/3}\left(  \frac{2}{3\sqrt{a}}\right)
,\ \ \nonumber\\
\int\limits_{-\infty}^{\infty}x\sin\left(  x+\frac{ax^{3}}{3}\right)  dx  &
=\frac{2}{\sqrt{3}a}K_{2/3}\left(  \frac{2}{3\sqrt{a}}\right)  . \label{eq11}%
\end{align}
Conserving the first (independent on $x)$ terms in Eq.(\ref{eq10}) we obtain
the probabilities in SQA%

\begin{align}
W_{i}^{(SQA)}  &  =-\frac{\mathrm{\operatorname{Im}}\kappa_{i}}{\omega}%
=\frac{\alpha m^{2}}{3\sqrt{3}\pi\omega}\int\limits_{-1}^{1}\frac{s_{i}%
}{1-v^{2}}K_{2/3}\left(  z\right)  dv,\ \ z=\frac{8}{3\left(  1-v^{2}\right)
\kappa},\ \nonumber\\
s_{2}  &  =2(3-v^{2}),\ \ \ s_{3}=3+v^{2},\ \ \ \kappa^{2}=4r(\mu^{2}+\nu
^{2})=-\frac{e^{2}}{m^{6}}\left(  F^{\mu\nu}k_{\nu}\right)  ^{2}. \label{eq12}%
\end{align}

The correction to SQA has a form%

\begin{equation}
W_{i}^{(1)}=-\frac{\alpha m^{2}\widetilde{\mathcal{F}}}{15\sqrt{3}\pi
\omega\kappa}\int\limits_{-1}^{1}\frac{dv}{1-v^{2}}G_{i}\left(  v,z\right)
,\ \ \widetilde{\mathcal{F}}=\frac{e^{2}\mathcal{F}}{m^{4}}=\frac{\nu^{2}%
-\mu^{2}}{2}, \label{eq13}%
\end{equation}
where%

\begin{align}
G_{2}\left(  v,z\right)   &  =\left(  36+4v^{2}-18z^{2}\right)  K_{1/3}\left(
z\right)  +\left(  3v^{2}-57\right)  zK_{2/3}\left(  z\right)  ,\nonumber\\
G_{3}\left(  v,z\right)   &  =-\left(  34+2v^{2}+36z^{2}\right)
K_{1/3}\left(  z\right)  +\left(  78-6v^{2}\right)  zK_{2/3}\left(  z\right)
. \label{eq14}%
\end{align}
The mathematical transformations of integrals can be found in Appendix C
\cite{[8]}. It is seen that in this order of decomposition the correction does
not depend on the invariant parameter $\mathcal{G}$, because $\mathcal{G}$ is
the pseudoscalar. The asimptotic of the integrals incoming in the correction
terms have been given in the mentioned Appendix C. The asymptotic at
$\kappa\ll1$ will become necessary further
\begin{equation}
W_{2}^{(1)}=\frac{4\alpha m^{2}\widetilde{\mathcal{F}}}{5\omega\kappa^{2}%
}\sqrt{\frac{2}{3}}\exp\left(  -\frac{8}{3\kappa}\right)  ,\ \ W_{3}%
^{(1)}=2W_{2}^{(1)},\ \ \frac{\ W_{i}^{(1)}}{W_{i}^{(SQA)}}=\frac
{64\widetilde{\mathcal{F}}}{15\kappa^{3}}. \label{eq15}%
\end{equation}

\section{\ \ Region of intermediate photon energies}

In the field which is weak comparing with the critical field $E/E_{0}=\nu\ll1$
( $E_{0}=1.32\cdot10^{16}%
\operatorname{V}%
/%
\operatorname{cm}%
$), $H/H_{0}$ = $\mu\ll1$ $(H_{0}=4.41\cdot10^{13}%
\operatorname{G}%
)$ and at the relatively low photon energies $r\lesssim\nu^{-2/3}$ the
standard quasiclassical approximation Eq.(\ref{eq12}) is non-applicable. This
follows from the last equality in Eq.(\ref{eq15}). For these energies, if the
condition $r\gg\nu^{2}$ is fulfilled, the method of stationary phase can be
applied at integration over $x$ in Eq.(\ref{eq4}). In this case the small
values of $v$ contribute to the integral over $v$. So one can expand the phase
$\psi(v,x)$ over $v$ and extend the integration limit to the infinity. We get%

\begin{equation}
\Omega_{i}=\frac{\alpha m^{2}}{2\pi\text{\textrm{i}}}\mu\nu\int
\limits_{-\infty}^{\infty}\ dv\int\limits_{-\infty}^{\infty\text{ }}%
\ f_{i}(0,x)\exp\left\{  -\text{\textrm{i}}\left[  \varphi\left(  x\right)
+v^{2}\chi\left(  x\right)  \right]  \right\}  xdx, \label{eq16}%
\end{equation}
where%

\begin{align}
\varphi\left(  x\right)   &  =2r\left(  \frac{1}{\mu}\tan\frac{\mu x}{2}%
-\frac{1}{\nu}\tanh\frac{\nu x}{2}\right)  +x,\ \nonumber\\
\ \chi\left(  x\right)   &  =rx^{2}\left(  \frac{\nu}{\sinh(\nu x)}-\frac{\mu
}{\sin(\mu x)}\right)  . \label{eq17}%
\end{align}
From the equation $\varphi^{\prime}(x_{0})=0$ we find the saddle point $x_{0}$%

\begin{equation}
\tan^{2}\frac{\nu s}{2}+\tanh^{2}\frac{\mu s}{2}=\frac{1}{r},\ \ x_{0}%
=-\mathrm{i}s.\label{eq18}%
\end{equation}
Substituting this value of $\U{445} _{0}$ in the expressions determined the
integrals in Eq.(\ref{eq16}) we have%

\begin{align}
\text{\textrm{i}}\varphi\left(  x_{0}\right)   &  =2r\left(  \frac{1}{\mu
}\tanh\frac{\mu s}{2}-\frac{1}{\nu}\tan\frac{\nu s}{2}\right)  +s\equiv
b(s),\ \label{eq19}\\
\ \text{\textrm{i}}\chi\left(  x_{0}\right)   &  =rs^{2}\left(  \frac{\nu
}{\sin(\nu s)}-\frac{\mu}{\sinh(\mu s)}\right)  \equiv\frac{1}{2}%
rs^{2}A(s),\label{eq191}\\
\ \text{\textrm{i}}\varphi^{\prime\prime}(x_{0})  &  =r\left[  \nu\sin
\frac{\nu s}{2}\diagup\cos^{3}\frac{\nu s}{2}+\mu\sinh\frac{\mu s}{2}%
\diagup\cosh^{3}\frac{\mu s}{2}\right]  \equiv rD\left(  s\right)  ,
\label{192}%
\end{align}

\begin{align}
f_{2}(0,x_{0})  &  =\frac{1}{\sinh(\mu s)\sin(\nu s)}\left[  \cos(\nu
s)\diagup\cosh^{2}\frac{\mu s}{2}-1\right]  \equiv-a_{2}(s),\ \nonumber\\
f_{3}(0,x_{0})  &  =\frac{1}{\sinh(\mu s)\sin(\nu s)}\left[  1-\cosh\mu
s\diagup\cos^{2}\frac{\nu s}{2}\right]  \equiv-a_{3}(s),\ \nonumber\\
\ f_{4}(0,x_{0})  &  =-\left(  4\cos^{2}\frac{\nu s}{2}\cosh^{2}\frac{\mu
s}{2}\right)  ^{-1}\equiv-a_{4}(s). \label{eq20}%
\end{align}

Performing the standard procedure of the stationary phase method and using
Eqs.(\ref{eq1})-(\ref{eq2}) one obtains the following expressions%

\begin{align}
\Omega_{i}  &  =a_{i}\frac{\alpha m^{2}\mu\nu}{r\sqrt{AB}}\exp\left(
-b\right)  ,\ \ W_{i}=\lambda_{i}\frac{\alpha m^{2}\mu\nu}{\omega\sqrt{AB}%
}\exp\left(  -b\right)  ;\label{eq21}\\
\lambda_{2}  &  =a_{2}-\frac{2a_{4}^{2}}{a},\ \ \lambda_{3}=a_{3}+\frac
{2a_{4}^{2}}{a},\ \ a=a_{3}-a_{2}+\sqrt{\left(  a_{3}-a_{2}\right)
^{2}+4a_{4}^{2}};\nonumber\\
b_{2}^{\mu}  &  =\ \left(  Bk\right)  ^{\mu}+\frac{2a_{4}}{a}\left(
Ck\right)  ^{\mu},\ \ b_{3}^{\mu}=\left(  Ck\right)  ^{\mu}-\frac{2a_{4}}%
{a}\left(  Bk\right)  ^{\mu} \label{eq22}%
\end{align}

These equations is valid at $r\gg1$ if the condition $b\gg1$ is fulfilled. The
first two terms of the decomposition of the functions $s\left(  r\right)  $
Eq.(\ref{eq18})) and $b\left(  s\left(  r\right)  \right)  $ Eq.(\ref{eq19})
over $1/r$ are%

\begin{equation}
s(r)\simeq\frac{4}{\kappa}\left(  1-\frac{8\widetilde{\mathcal{F}}}%
{3\kappa^{2}}\right)  ,\ \ b(r)\simeq\frac{8}{3\kappa}-\frac{64\widetilde
{\mathcal{F}}}{15\kappa^{3}},\ \ \kappa^{2}=4(\mu^{2}+\nu^{2})r. \label{eq23}%
\end{equation}
It is follows from this formula that the applicability of Eq.(\ref{eq21}) is
limited by the condition $\kappa\ll1$. The main values of the rest terms in
Eqs.(\ref{eq21}),(\ref{eq22}) have a form%

\begin{align}
A  &  =\frac{1}{3}\left(  \mu^{2}+\nu^{2}\right)  s,\ D=\frac{3}{2}%
A;\ a_{2}=\frac{\mu^{2}+2\nu^{2}}{4\mu\nu},\ a_{3}=\frac{2\mu^{2}+\nu^{2}%
}{4\mu\nu},\nonumber\\
\ \ \ a_{4}  &  =\frac{1}{4},\ \ a=\frac{\mu}{2\nu},\ \ \lambda_{2}=\frac
{\mu^{2}+\nu^{2}}{4\mu\nu},\ \ \lambda_{3}=2\lambda_{2},\ \label{eq24}%
\end{align}
and the vectors of polarization are given by Eq.(\ref{eq8}). Substituting this
values into equation for $W_{i}$ we have%

\begin{equation}
W_{2}=\frac{\alpha m^{2}\kappa}{8\omega}\sqrt{\frac{3}{2}}\exp\left(
-\frac{8}{3\kappa}+\frac{64\widetilde{\mathcal{F}}}{15\kappa^{3}}\right)
,\ \ W_{3}=2W_{2}. \label{eq25}%
\end{equation}
In the region of the SQA applicability and for $\kappa\ll1$ this probability
coincides with the results of the previous section and so the overlapping
region of both approximations exists.

It is interesting to consider the photon energy region $|r-1|\ \ll1$ in the
presence of a weak electric field $(\nu\ll\mu)$ where in the absence of an
electric field the approach under consideration is valid if the condition
$r-1\gg\mu$ is fulfilled \cite{[8]}. In this case Eq.(\ref{eq18}) and its
solutions are given by the following approximate equations%
\begin{align}
\frac{\xi^{2}y_{0}^{2}}{16}  &  \simeq\exp(-y_{0})+\frac{1-r}{4},\ \ y_{0}=\mu
s,\ \xi=\frac{\nu}{\mu};\label{eq251}\\
\ y_{0}  &  \simeq2\ln\frac{2}{\xi\ln\frac{4}{\xi}}\left(  1-\frac{r-1}%
{2\xi^{2}\ln\frac{2}{\xi}\ln^{3}\frac{4}{\xi}}\right)  ,\ |r-1|\ \lesssim
\xi^{2};\\
y_{0}  &  \simeq\ln\frac{4}{r-1}\left(  1-\frac{\xi^{2}}{4\left(  r-1\right)
}\ln\frac{4}{r-1}\right)  ,\ \ r-1\gg\xi^{2};\\
\xi y_{0}  &  =\nu s\simeq2\sqrt{1-r},\ \ 1-r\gg\xi^{2}.
\end{align}
The applicability of the using saddle-point method is connected with the large
value of the coefficient to the second power $(y-y_{0})^{2}$ of the
decomposition in the phase Eq.(\ref{eq17}). In the energy region under
consideration we have
\begin{equation}
\mathrm{i}\varphi^{\prime\prime}(x_{0})(x-x_{0})^{2}/2\simeq\frac{\xi^{2}%
}{4\mu}\left[  y_{0}+\frac{y_{0}^{2}}{2}+\frac{2\left(  r-1\right)  }{\xi^{2}%
}\right]  (y-y_{0})^{2}. \label{eq252}%
\end{equation}
So, we have from the upper equations that in the case $\nu/\mu=\xi
\ll1,\ |r-1|\ \lesssim\xi^{2}$ Eq.(\ref{eq21}) is valid if the condition
$\xi^{2}/\mu\gg1$ is fulfilled. In the case $1\gg r-1\ \gg\xi^{2}\ $the
condition $r-1\ \gg\mu$ has to be available for that. And in the case
$1\gg1-r$ $\gg\xi^{2}$ the condition $\sqrt{1-r}\xi/\mu=\sqrt{\left(  \xi
^{2}/\mu\right)  \left(  1-r\right)  /\mu}\gg1$ is necessary for the
applicability of the approach under consideration.

At low photon energy $r\ll1\ $($\nu^{2}\ll r\ll\nu^{2/3})$ we have%

\begin{align}
\nu s  &  \simeq\pi-2\sqrt{r}+r^{3/2}\left(  \frac{2}{3}-\tanh^{2}\frac
{\pi\eta}{2}\right)  ,\nonumber\\
b  &  \simeq\frac{1}{\nu}\left(  \pi-4\sqrt{r}+\frac{2r}{\eta}\tanh\frac
{\pi\eta}{2}\right)  ;\label{eq26}\\
a_{2}  &  =\frac{1}{\sqrt{r}\sinh(\pi\eta)}\left(  1-\frac{1}{2}\tanh^{2}%
\frac{\eta\pi}{2}+\frac{\mu}{4r}\coth\pi\eta\right)  ,\ \nonumber\\
\ a_{3}  &  =\frac{\coth(\pi\eta)}{2r^{3/2}}\left(  1+\frac{4\eta\sqrt{r}%
}{\sinh(2\pi\eta)}\right)  \simeq a,\ \ a_{4}=\left(  4r\cosh^{2}\frac{\eta
\pi}{2}\right)  ^{-1},\label{eq27}\\
\ \ \lambda_{2}  &  =\frac{1}{\sqrt{r}\sinh(\pi\eta)}\left[  1-\left(
\frac{1}{2}+\frac{1}{\cosh(\pi\eta)}\right)  \tanh^{2}\frac{\eta\pi}{2}%
+\frac{\mu}{4r}\coth(\pi\eta)\right]  ,\nonumber\\
\lambda_{3}  &  \simeq a_{3},\ \ A=\frac{\nu}{\sqrt{r}}\left(  1-\frac
{2\eta\sqrt{r}}{\sinh(\pi\eta)}\right)  ,\ \ D=\frac{\nu}{r^{3/2}}%
,\ \eta=\frac{\mu}{\nu}. \label{eq28}%
\end{align}
Here we have retained the leading and the leading to leading terms of
decomposition. The term $\propto\mu$ in $a_{2}$ has appeared as the
contribution of the second term in $f_{1}$ ($\propto v^{2}$) in Eq.(\ref{eq5}%
). Substituting these values into Eq.(\ref{eq21}) one obtains the following
expression for the probability of the process%

\begin{align}
W_{3}  &  =\frac{\alpha m^{2}\mu}{2\omega\sqrt{r}}\coth\left(  \pi\eta\right)
\left(  1+\frac{\eta\sqrt{r}}{\sinh(\pi\eta)}+\frac{4\eta\sqrt{r}}{\sinh
(2\pi\eta)}\right)  \exp\left(  -b\right)  ,\nonumber\\
W_{2}  &  =\frac{\alpha m^{2}\mu\sqrt{r}}{\omega\sinh(\pi\eta)}\left[
1-\frac{2+\cosh(\pi\eta)}{2\cosh(\pi\eta)}\tanh^{2}\frac{\eta\pi}{2}+\frac
{\mu}{4r}\coth(\pi\eta)\right]  \exp\left(  -b\right)  , \label{eq29}%
\end{align}
where $b$ is given by Eq.(\ref{eq26}). One can see out of this equation that
$W_{2}\ll W_{3}.$At $\eta\gg1$ the probability $W_{3}$ has been increased by
the factor $\eta\pi\exp\left(  \pi r/\nu\right)  $ in comparison with the case
of the absence of magnetic field. The probability $W_{2}$ has been reduced by
the additional factor $(\exp\left(  -\pi\mu/\nu\right)  )$ and becomes
non-applicable at $\mu\gtrsim\sqrt{r}\gg\nu$. In that case for the probability
$W_{2}$ one can use Eq.(\ref{eq32}) which will be get below.

\section{\ \ Approximation at low photon energy}

At $r\sim\nu^{2}$ the above approximation becomes non-applicable and another
approach has to be. We close the integration over $x$ contour in
Eq.(\ref{eq4}) in the lower half-plane and represent this equation in the
following form%

\begin{equation}
\Omega_{i}=\frac{\alpha m^{2}}{2\pi\text{\textrm{i}}}\mu\nu\int\limits_{-1}%
^{1}\ dv\sum\limits_{n=1}^{\infty}\oint f_{i}(v,x)\exp(\text{\textrm{i}}%
\psi(v,x))xdx, \label{eq30}%
\end{equation}
where the path of integration is any simple closed contour around the point
$-$\textrm{i}$\pi n\diagup\nu.$ Let us choose the contour near this point in
the following way $\nu x=-$\textrm{i}$\pi n+\xi_{n},\ \ |\xi_{n}|\ \sim
\sqrt{r}\sim\nu$ and expand the function entering in over the variables
$\xi_{n}$. In the case $\nu\ll1,$ because of appearance of the factor
$\exp\left(  -\mathrm{i}\pi n\diagup\nu\right)  ,$\ the main contribution to
the sum gives the term $n=1$. Near the point $-$\textrm{i}$\pi\diagup\nu$ the
main terms of expansion such as $(\xi\equiv\xi_{1})$%

\begin{align}
f_{3}  &  =\frac{4\text{\textrm{i}}}{\xi^{3}}\coth(\pi\eta)\cos^{2}\frac{\pi
v}{2},\ \ f_{2}=-\frac{1}{\xi^{2}}\frac{\coth(\pi\eta)\text{ }}{\sinh(\pi
\eta)}\sinh(v\pi\eta)\ \sin(v\pi),\ \nonumber\\
\ f_{4}  &  =\frac{2}{\xi^{2}}\frac{\cosh(\pi\eta)-\cosh(v\pi\eta)}{\sinh
^{2}(\pi\eta)}\cos^{2}\frac{\pi v}{2},\ \ \psi=\frac{4r}{\xi\nu}\cos^{2}%
\frac{\pi v}{2}-\frac{\xi}{\nu}+\frac{\mathrm{i}\pi}{\nu}. \label{eq31}%
\end{align}
Using the integrals Eq.(7.3.1) and Eq.(7.7.1 (11)) in \cite{[15]} and
substituting the result in Eqs.(\ref{eq1})-(\ref{eq2}) we find%

\begin{align}
W_{3}  &  =2\frac{\alpha m^{2}}{\omega}\eta\pi\coth(\pi\eta)\text{ }%
\exp\left(  -\frac{\pi}{\nu}\right)  I_{1}^{2}\left(  z\right)  ,\ \ \ z=\frac
{2\sqrt{r}}{\nu},\label{eq311}\\
W_{2}  &  =\frac{\alpha m^{2}}{\omega}\mu\coth(\pi\eta)\exp\left(  -\frac{\pi
}{\nu}\right)  \left[  \frac{\pi\eta}{\sinh(\pi\eta)}\int\limits_{0}^{1}%
\cosh(v\pi\eta)I_{0}\left(  2z\cos\frac{\pi v}{2}\right)  dv-1\right]  ,
\label{eq32}%
\end{align}
where $\mathrm{I}_{n}\left(  z\right)  $ is the Bessel function of imaginary
argument. At calculation $W_{2}$ the integration by parts over $v$ has been
performed. For $\eta\ll1$ one obtains%
\begin{equation}
W_{2}=\frac{\alpha m^{2}}{\omega}\frac{\nu}{\pi}\exp\left(  -\frac{\pi}{\nu
}\right)  \left(  I_{0}^{2}\left(  z\right)  -1\right)  .
\end{equation}
The found probability is applicable for $r\ll\nu.$ Here we have kept the main
terms in $W_{i}$ only.

For $r\gg\nu^{2}$ the asymptotic representation $\mathrm{I}_{n}\left(
z\right)  \simeq\exp\left(  z\right)  \diagup\sqrt{2\pi z}$ can be used. As a
result one obtains the probability Eq.(\ref{eq29}) where the leading terms
have to be retained. At very low photon energy $r\ll\nu^{2},$ using the
expansion of the Bessel functions for the small value of argument, we have%

\begin{equation}
W_{3}=2\frac{\alpha m^{2}r}{\omega\nu^{2}}\eta\pi\coth(\pi\eta)\exp\left(
-\frac{\pi}{\nu}\right)  ,\ \ W_{2}=\frac{\nu}{\pi\left(  1+\eta^{2}\right)
}W_{3}. \label{eq33}%
\end{equation}

The probability under consideration is of interest of theoretics for arbitrary
values $\mu$ and $\nu$. For $r\ll\nu^{2}\diagup\left(  1+\nu^{2}\right)  $ one
can conserve in the phase $\psi(v,x)$ the term $-x$ only. After integrating
over $v$ we get the following equation for the probability averaged over the
photon polarizations%

\begin{align}
W  &  =\frac{W_{2}+W_{3}}{2}=\frac{\alpha m^{2}r}{\mathrm{i}\pi\omega}%
\sum\limits_{n=1}^{\infty}\oint F(y_{n})\exp\left(  -\mathrm{i}\frac{y_{n}%
}{\nu}\right)  dy_{n},\nonumber\\
F(y)  &  =\frac{\cosh(y)\left(  \eta y\cos\left(  \eta y\right)  -\sin\left(
\eta y\right)  \right)  }{\sinh y\sin^{3}\eta y}+\frac{\eta\cos(\eta y)\left(
y\cosh y-\sinh y\right)  }{\sinh^{3}y\sin(\eta y)}. \label{eq34}%
\end{align}
Summing the residues in the points $y_{n}=-\mathrm{i}n\pi$ one obtains%

\begin{align}
W  &  =\frac{\alpha m^{2}r}{\omega}\sum\limits_{n=1}^{\infty}\exp\left(
-\frac{\pi n}{\nu}\right)  \Phi\left(  z_{n}\right)  ,\ \ z_{n}=\eta\pi
n,\label{eq35}\\
\Phi\left(  z_{n}\right)   &  =\frac{z_{n}}{\nu^{2}}\coth z_{n}+\frac{2}%
{\sinh^{2}z_{n}}\left[  \frac{\eta z_{n}}{\nu}+\left(  1+\eta^{2}\right)
z_{n}\coth z_{n}-1\right]  . \label{eq36}%
\end{align}
In the absence of magnetic field ( $\eta\rightarrow0,$ $z_{n}\rightarrow0$ )
we have%

\begin{align}
\Phi &  =\frac{1}{\nu^{2}}+\frac{2}{\nu\pi n}+\frac{2}{\pi^{2}n^{2}}+\frac
{2}{3},\nonumber\\
W  &  =\frac{\alpha m^{2}r}{\omega}\left[  \left(  \frac{1}{\nu^{2}}+\frac
{2}{3}\right)  \frac{1}{e^{\pi/\nu}-1}-\frac{2}{\pi\nu}\ln\left(
1-e^{-\pi/\nu}\right)  +\frac{2}{\pi^{2}}\mathrm{Li}_{2}\left(  e^{-\pi/\nu
}\right)  \right]  , \label{eq37}%
\end{align}
where $\mathrm{Li}_{2}\left(  z\right)  $ is the Euler dilogarithm. In the
opposite case $\eta\gg1$ one obtains%

\begin{equation}
\Phi=\frac{\pi\eta n}{\nu^{2}},\ \ W=\frac{\alpha m^{2}r}{\omega\nu^{2}}%
\frac{\pi\eta}{4}\sinh^{-2}\frac{\pi}{2\nu}. \label{eq38}%
\end{equation}

\section{Conclusion}

We have considered the process of pair creation in constant and homogeneous
electromagnetic fields with a real photon taking part in. The probability of
the process has been calculated using three different overlapping
approximation. In the region of SQA applicability the created by a photon
particles have ultrarelativistic energies. The role of fields in this case is
to transfer the required transverse momentum and the electric and magnetic
field actions are equivalent. But even in this case it is necessary to note a
special significance of a weak electric field $E=\xi H$ $(\xi\ll1)$ in the
removal of the root divergence of the probability when the particles of pair
are created on the Landau levels with the electron and positron momentum
$p_{3}=0$ \cite{[8]}. The frame is used where $k_{3}=0.$

Generally speaking, at $\xi\ll1$ the formation time $t_{c}$ of the process
under consideration is $1/\mu.$ Here we use units $\hbar=c=m=1.$ At this time
the particles of creating pair gets the momentum $\delta p_{3}\sim\xi.$ If the
value $\xi^{2}$ becomes more larger than the distance apart Landau levels
$2\mu$ $(\nu^{2}\gg\mu^{3})$ all levels have been overlapped. Under this
condition the divergence of the probability is vanished and the new
quasiclassical approach is valid even in the energy region $r-1\lesssim\mu$
where it has been inapplicable in the absence of electric field \cite{[8]}. In
the opposite case $\nu^{2}\ll\mu^{3}$ for the small value of $p_{3}\ll
\sqrt{\mu},$ in the region where the influence of electric field is
negligible, the formation time of the process $t_{f}$ is $1/p_{3}^{2}$ and
$\delta p_{3}\sim\nu/p_{3}^{2}$ $\ll p_{3}$ .\ It is follows from above that
$\nu^{1/3}\ll p_{3}\ll\sqrt{\mu}$ . At this condition the value of
discontinuity is $\sqrt{t_{f}/t_{c}}\sim\sqrt{\mu}/p_{3}.$For $\nu^{1/3}\gg
p_{3}$ the time $t_{f}$ is determined by the self-consistent equation
$\delta\varepsilon^{2}\sim1/t_{f}\sim\nu^{2}t_{f}^{2},$ $t_{f}\sim\nu^{-2/3}$
and the value of discontinuity becomes $\sqrt{\mu t_{f}}\sim(\mu^{3}/\nu
^{2})^{1/6}$ instead of $\sqrt{\mu}/p_{3}.$

In the region $\omega\lesssim2m$ $(r\lesssim1)$ the energy transfer from
electric field to the created particles becomes appreciable and for $\omega
\ll$ $m$ it determines the probability of the process mainly. At $\omega\ll
eE/m$ the photon assistance in the pair creation comes to the end and the
probability under consideration defines the probability of photon absorption
by the particles created by electromagnetic fields. The influence of a
magnetic field on the process is connected with the interaction of the
magnetic moment of the created particles and magnetic field. This interaction,
in particular, has appeared in the distinction of the pair creation
probability by field for scalar and spinor particles \cite{[5]}.

\end{document}